\documentstyle[epsfig]{elsart}
\begin{document}
\begin{frontmatter}
\title{Design, construction, and operation of SciFi tracking detector for K2K 
       experiment}
\collab{K2K Collaboration}
\author[Kobe]{A. Suzuki}
\author[KEK]{H. Park}
\author[Kobe]{S. Aoki}
\author[Kobe]{S. Echigo}
\author[Kobe]{K. Fujii}
\author[Kobe]{T. Hara}
\author[Kobe]{T. Iwashita}
\author[Kobe]{M. Kitamura}
\author[Kobe]{M. Kohama}
\author[Kobe]{G. Kume}
% ...

\address[Kobe]{Kobe University, Kobe, Japan}
\address[Kyoto]{Kyoto University, Kyoto, Japan}
\address[KEK]{High Energy Accelerator Research Organization(KEK), Tsukuba, Japan}
\address[ICRR]{Institute for Cosmic Ray Research, University of Tokyo, Tanashi, Japan} 
\address[Niigata]{Niigata University, Ikarashi, Niigata, Japan}
\address[Okayama]{Okayama University, Okayama, Japan}
\address[Osaka]{Osaka University, Toyonaka, Osaka, Japan}
\address[SUT]{Science University of Tokyo, Noda, Chiba, Japan}
\address[Tohoku]{Tohoku University, Sendai, Miyagi, Japan}
\address[Tokai]{Tokai University, Kanagawa, Japan}
\address[SNU]{Seoul National University, Seoul, Korea}
\address[CNU]{Chonnam National University, Kwangju, Korea}
\address[Dongshin]{Dongshin University, Naju, Korea}
\address[BU]{Boston University, Boston, USA}
\address[UCI]{University of California, Irvine, USA}
\address[Hawaii]{University of Hawaii, USA}
\address[SUNY]{State University of New York, Stony Brook, USA}
\address[UW]{University of Washington, Seatle, USA}
\address[Warsaw]{Warsaw University, Poland}
\thanks[Chikamatsu]{Present address: Miyagi Women's College, Sendai, 
Miyagi, Japan}
\begin{abstract}

     We describe the construction and performance of a scintillating fiber
detector used in the near detector for the K2K (KEK to Kamioka, KEK E362)
long baseline neutrino oscillation experiment. The detector uses 3.7 m long
and 0.692 mm diameter scintillating fiber coupled to image-intensifier tubes
(IIT), and a CCD camera readout system. Fiber sheet production and detector
construction began in 1997, and the detector was commissioned in March, 1999. 
Results from the first K2K runs confirm good initial performance : position
resolution is estimated to be about 0.8 mm, and track finding efficiency is
$98 \pm 2$ \% for long tracks (i.e., those which intersect more than 5
fiber planes).  The hit efficiency was estimated to be $92 \pm 2$ \%
using cosmic-ray muons, after noise reduction at the offline stage.  The
possibility of using the detector for particle identification is also
discussed.
 
\end{abstract}
\end{frontmatter}
\section{Introduction}
\subsection{K2K experiment}
\hspace{3mm} The Super-Kamiokande Collaboration has shown strong evidence for
neutrino oscillations in an analysis of atmospheric neutrino data\cite{SK}.

\hspace{3mm} K2K\cite{K2K} is a long baseline neutrino experiment designed to
allow more precise studies of neutrino oscillations. A high-purity muon
neutrino beam, generated using the KEK 12 GeV proton synchrotron, is directed
through a near detector system at KEK to the Super-Kamiokande site 250 km
away. By comparing the $\nu_{\mu}$ beam flux and energy spectrum between the
near and far detectors, we can investigate neutrino oscillations between
$\nu_{\mu}$ and other neutrino flavors.

\hspace{3mm} The near detector includes a 1kt water Cherenkov detector and a
fine-grained detector (FGD), which in turn consists of the scintillating
fiber (SciFi) detector, scintillating counters, a lead glass counter, and a
muon range detector. Data taking for K2K began in March, 1999.

\subsection{Scintillating fiber (SciFi) detector}
\hspace{3mm} The near neutrino detector is designed to measure the flux and
the energy spectrum of the neutrino beam as it leaves KEK. The detector must
provide good tracking capability, allowing discrimination between different
types of interactions such as quasi-elastic or inelastic events. Its mass
composition should be dominated by water, to allow cancellation of common
systematic uncertainties with the Super-Kamiokande detector to the maximum
extent possible.  In order to study neutrino interactions in greater detail,
we decided to use a SciFi detector constructed with fiber tracking layers
interleaving water target tanks. SciFi detectors have been used in UA2 and
CHORUS experiments \cite{UA2,CHORUS}.  The K2K SciFi detector is the largest
ever so far, and employs a simple design of optoelectronics readout and a
calibration system using electro-luminescent (EL) plates.

\section{Construction and operation of the K2K SciFi detector}
\subsection{Design and structure}
\begin{figure}[htbp]
  \begin{center}
    \epsfig{file=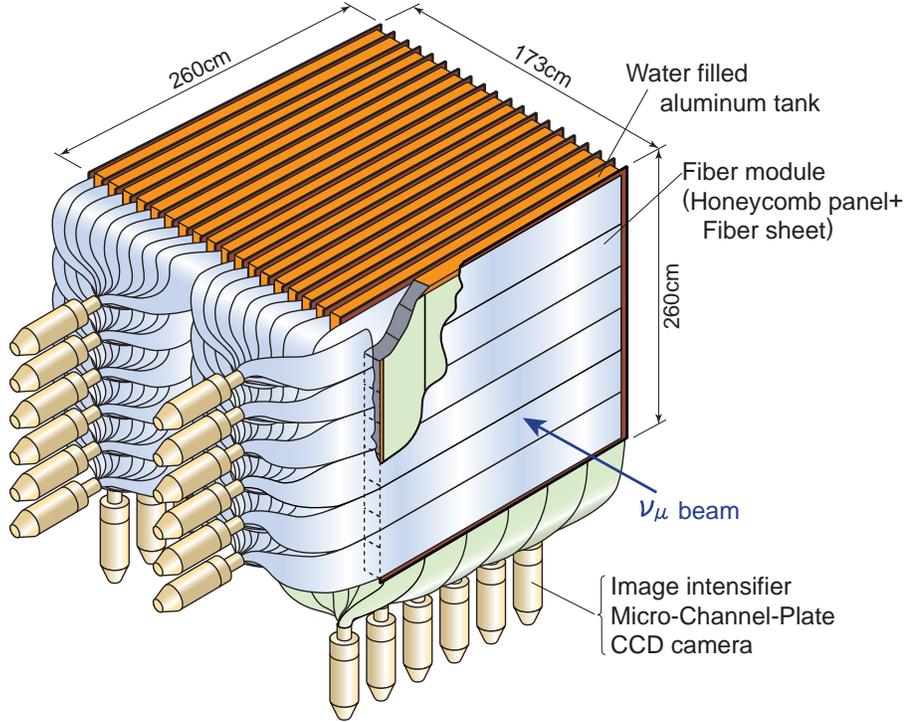,width=12cm}
    \caption{Schematic overview of the SciFi tracking detector.}
    \label{fig:scifi_birdeye}
  \end{center}
\end{figure}

\begin{figure}[h]
  \begin{center}
    \epsfig{file=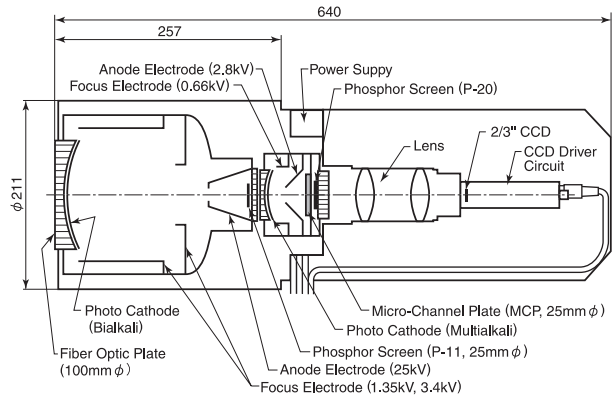,width=8cm}
    \caption{IIT-CCD chain.}
    \label{fig:IIT-CCD}
  \end{center}
\end{figure}

\hspace{3mm} A schematic overview of the SciFi detector is shown in 
Fig. \ref{fig:scifi_birdeye}.
It consists of 20 layers of 2.6 m $\times$ 2.6 m
tracking modules, spaced 9 cm apart, each of which contains double layers of
scintillating fiber sheets in both the horizontal and vertical directions
(i.e., XXYY layers). 
The fiber sheets are coupled to image-intensifier tubes
(IITs) which in turn are read out by CCD cameras (Fig. \ref{fig:IIT-CCD}).
Between the fiber modules, there are 19 layers of water target contained in
extruded aluminum tanks. Each target layer consists of 15 tank modules, and
thus there are 285 in total. The width, height, and length of the tanks are 6
cm, 16 cm, and 240 cm, respectively. The thickness of the aluminum tank is
1.8 mm.  The fiber is 3.7 m long and 0.692 mm in diameter. 

\subsection{Optimization of the scintillating fibers}
\subsubsection{Measurement of the light yield of various fibers}
\hspace{3mm} We measured the light yield and attenuation length of various
types of scintillating fibers using a photo-multiplier tube (PMT),
irradiating the fiber with a $^{90}$Sr $\beta$-ray source.  The light yield,
after propagation of the light over a distance $x$ in the fiber, is expressed
by
\begin{eqnarray}
Y(x) = Y_0 \exp[-x/\lambda],
\end{eqnarray}
where $Y_0$ is an initial light yield at $x=0$ (IIT surface) and $\lambda$ is
an attenuation length. We estimated $Y_0$ by extrapolation from measurements
at $x$ = 1.5, 2.0, 2.5, and 3.0 m. The results are summarized in Table
\ref{tab:scifi_efficiency}.
\begin{table}[b]
\caption{Measured light yield $Y_{0}$ and attenuation length $\lambda$.  The
detection efficiency $\varepsilon$ after 4 m light propagation for a
double-layered fiber sheet was calculated using $Y_{0}$ and $\lambda$.}
\label{tab:scifi_efficiency}
\begin{tabular}{lcccc} 
SciFi type & Diameter[mm] & $Y_{0}$[p.e.] & $\lambda$[m] & 
Efficiency$(\varepsilon$)[\%] \\
\hline
SCSF-78M & 0.5 & 4.4 $\pm$ 0.2 & 3.46 $\pm$ 0.18 & 97.2 $\pm$ 0.7 \\
         & 0.7 & 6.5 $\pm$ 0.2 & 3.57 $\pm$ 0.17 & 99.6 $\pm$ 0.1 \\
SCSF-78  & 0.7 & 5.2 $\pm$ 0.1 & 3.48 $\pm$ 0.13 & 98.6 $\pm$ 0.3 \\
         & 0.9 & 7.2 $\pm$ 0.2 & 3.86 $\pm$ 0.13 & 99.9 $\pm$ 0.03 \\
SCSF-77  & 0.7 & 4.2 $\pm$ 0.1 & 4.39 $\pm$ 0.27 & 98.9 $\pm$ 0.3 \\
BCF-12   & 0.7 & 3.6 $\pm$ 0.1 & 3.36 $\pm$ 0.18 & 94.2 $\pm$ 1.2 \\
\hline
\end{tabular}
\end{table}
Kuraray SCSF-78M is a multi-cladding fiber, and the others have
single-cladding. From these measurements, we also estimated the detection
efficiency for minimum ionizing particles after light propagation over a
distance $x$=4 m for a double-layered fiber sheet. Here, we took into account
the quantum efficiencies of the IIT ($22 \%$) and PMT ($25 \%$), and the
reflectivity at the far end of the fiber ($70 \%$). When we required more
than 99 \% detection efficiency, there were only two candidates left. In
order to minimize the readout area, we selected the smaller-diameter fiber,
SCSF-78M (0.7 mm diameter).

\subsubsection{Aging}
\begin{figure}[ht]
  \begin{center}
    \epsfig{file=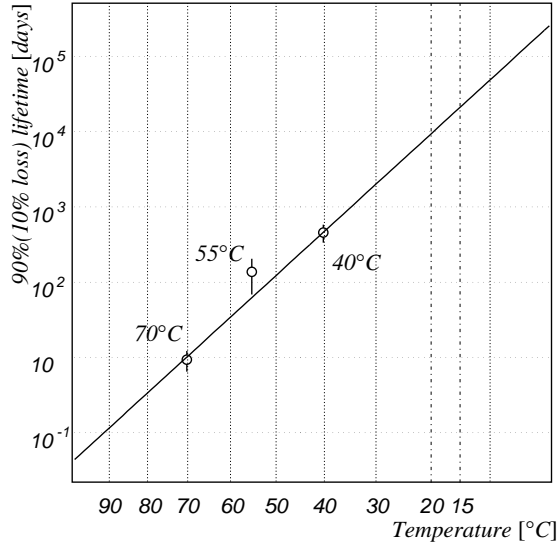,width=9cm}
    \caption{Fiber lifetime as a function of temperature. The measured fiber 
	was Kuraray SCSF-78M version 1.1(Anti-O$_{2}$ type).}
    \label{fig:Arrhenius}
  \end{center}
\end{figure}

\hspace{3mm} Scintillating fibers are known to have a finite lifetime, with
fiber transparency diminishing with age.  Aging is mainly caused by a
chemical reaction of the core material (polystyrene) with oxygen in
air. Fiber lifetime depends on temperature. We measured the decrease in light
yield of SCSF-78M at several elevated temperatures and estimated the fiber
lifetime at lower temperatures using an ``Arrhenius'' plot, lifetime as a
function of temperature (Fig. \ref{fig:Arrhenius}).
Here we define a fiber lifetime as the number of days at which the light
yield of the fiber drops to 90 \% of its initial value. From the Arrhenius
plot, the lifetimes of SCSF-78M at 15 $^{\circ}$C and 20 $^{\circ}$C are
estimated to be 21,000 and 9,400 days, respectively. If we keep the
temperature of the experimental hall less than 20 $^{\circ}$C, the decrease
in the light yield is expected to be less than 10 \% during the five years of
the experiment.  In fact, we attempt to maintain the SciFi detector
temperature at 16 $^{\circ}$C.

\subsection{Fiber sheet production}
\begin{figure}[htbp]
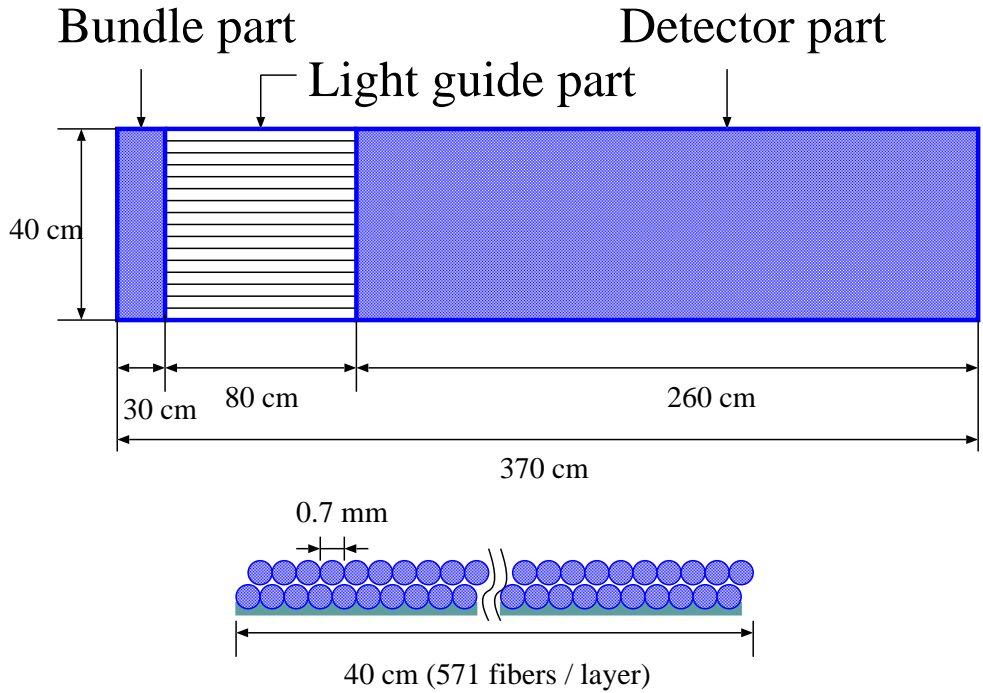

  \begin{center}
    \epsfig{file=figs/sover.eps,width=13cm}
    \epsfig{file=figs/scross.eps,width=7cm}
    \caption{Schematic view of a SciFi sheet. The upper figure shows a 
    top view and the lower figure shows a cross-sectional view.}
    \label{fig:SCIFI_sheet}
  \end{center}
\end{figure}

\hspace{3mm} We had to fabricate double-layered fiber sheets before building
the fiber tracking modules. Each fiber sheet is 370 cm long and 40 cm
wide, consisting of a 260 cm long sensitive area, 80 cm long light guide, and
30 cm long IIT bundle segment (Fig. \ref{fig:SCIFI_sheet}).

\hspace{3mm} A fiber tracking module consists of 6 double-layered fiber
sheets laid side by side to make a total width of 240 cm, in both horizontal
and vertical directions. For 20 such modules, we needed to fabricate 240
sheets in total, plus 8 spares. This method of fiber sheet production is
quite similar that used by the CHORUS Collaboration \cite{nagoya-1,nagoya-2}. 
To produce 3.7 m long fiber sheets, we prepared a 1.2 m diameter drum with a
spiral groove whose pitch was 0.7 mm. We wound fiber stock along the groove
until the sheet became 40 cm wide. Then, we coated the sheet with white
paint. After it became dry, we wound the second layer fiber onto the first
layer and painted it again. Finally, we cut the sheet, removed it from the
drum, and painted its inner side.

\begin{figure}[htbp]
  \begin{center}
    \epsfig{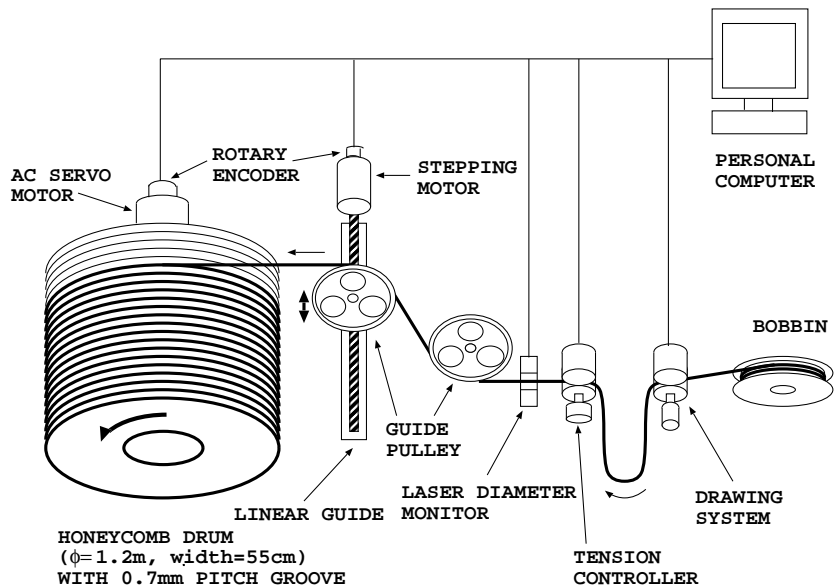}
    \vspace{10pt}
    \caption{Schematic view of the fiber sheet production system.}
    \label{fig:spool}
  \end{center}
\end{figure}

\hspace{3mm} A control system for fiber sheet production is shown in
Fig. \ref{fig:spool}.
Fiber stock was wound onto the bobbin and threaded onto the winding
machine. The drawing system, which consists of 2 rubber coated rollers, a
rotary encoder, and a stepping motor, drew fiber from the bobbin. Because the
two layers have slightly different circumferences on the drum, we applied a
constant tension to each layer (150 gw to the first and 300 gw to the second
layers, respectively) during winding. After being removed from the drum, the
prestressed sheet then became flat due to the restoring force supplied by the
outer fibers. At the downstream end of the tension controller, a laser sensor
monitored the fiber diameter. When it detected a fiber section out of
tolerance (692 $\mu$m $\pm 18 \mu$m), the whole system stopped, and we cut
away the segment. Guide pullies attached to an accurate linear guide, driven
by a stepping motor, lead the fiber onto the groove of the drum. They were
synchronized with rotation of the drum and moved 0.7mm for a turn of
rotation. The drum was driven by an AC servo motor. The rotation speed was
adjustable and the maximum speed was about 1 m/sec (0.265 rps). The whole
fiber sheet production system was controlled by a conventional PC with 2
custom interface boards.

\subsection{Fiber sheet modules}
\label{sec-quality}
\hspace{3mm} After all the fiber sheets were made, both ends of the sheets
were polished by a milling machine and far ends of the sheets were sputtered
with aluminum. We checked the quality of all the individual fibers by
irradiating the sheets with $^{90}$Sr $\beta$-source at the farthest
point. The light yield of the fiber sheets was measured to be 8$\pm$1
photo-electrons.
 
\hspace{3mm} Since the fiber sheets are very fragile, we needed a strong and
light material to support the fiber sheet. We chose a honeycomb panel with a
dimension of 2.6 m $\times$ 2.6 m $\times$ 1.6 cm, consisting of a 1.4 cm
thick paper-honeycomb core and two 1 mm thick GFRP skins. The area density of
the honeycomb board is 0.52 g/cm$^2$, and their net mass is 30 kg per board.

\hspace{3mm} We glued 6 fiber sheets side by side on one surface of the
honeycomb panel in a horizontal direction, and 6 more sheets on the other
side in a vertical direction (referring to the final orientation), using
epoxy adhesive CY221/HY2967. Since we found that gas released by the glue
actually accelerates fiber aging, we placed the glue in a vacuum chamber for
5 minutes before gluing. The positions and the straightness of the sheets
were controlled within 0.5 mm by a ruler bar and alignment pins when the
sheets are glued. The straightness of each sheet was measured by a micrometer
every 10 cm with accuracy of 0.1 mm.

\hspace{3mm} Since the fiber sheet modules were rather large (3.7 m $\times$
3.7 m), we made a special aluminum jig (4 m $\times$ 4 m) to perform
assembly, transportation, and installation of the modules safely and easily.
Throughout the installation, only about 20 fibers were broken out of 274080.

\subsection{Fiber bundles}
\hspace{3mm} In order to make effective use of the photosensitive area~(10~cm
diameter) of the IIT surface, fibers are bundled at the readout. All the
fibers (1142 $\times$ 6 $\times$ 20 $\times$ 2 = 274080) are grouped at the
readout end and glued into bundles, which are then attached to the 24
IIT's. Ten consecutive fiber sheets along the beam direction are assigned to
one fiber bundle. Thus, there are 6 fiber bundles in the upstream half of the
detector, and another 6 fiber bundles in the downstream half, in both x
(horizontal) and y (vertical) directions.

\begin{figure}[htbp]
  \begin{center}
    \epsfig{file=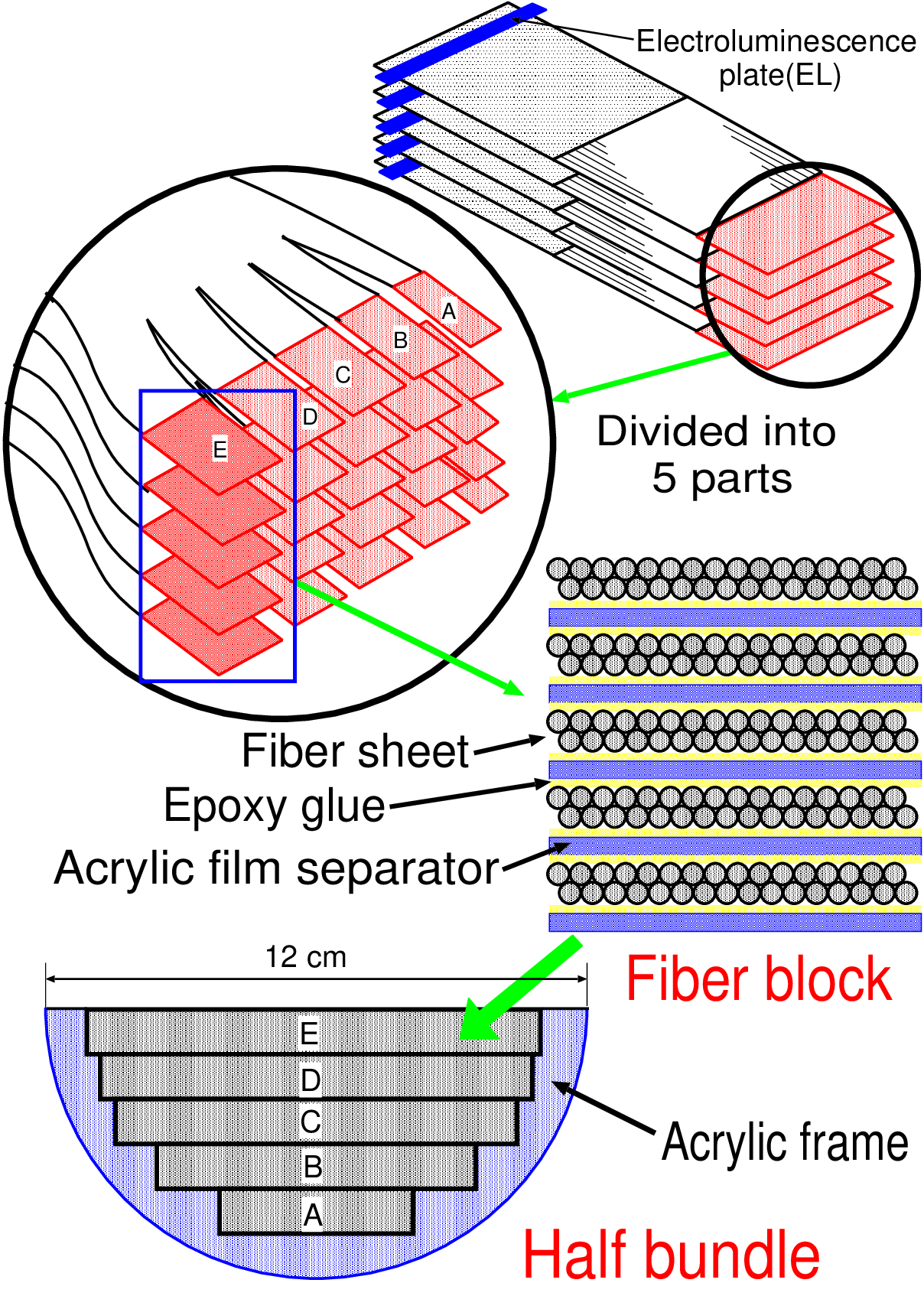,width=9cm}
    \caption{Fiber bundle assembly. Five fiber sheets are gathered into	
     each bundle.} 
    \label{fig:bundle}
  \end{center}
\end{figure}

\hspace{3mm} One fiber bundle contains 1142 $\times$ 10~(or 274080/24)= 11420
fibers. Fig. \ref{fig:bundle} shows the arrangement used to accumulate fibers
and make a (half) bundle from (5) fiber sheets.  The remaining half bundle is
symmetrical with respect to the direction along the SciFi layer and the two
halves are glued together.
In order to optically separate fiber sheets from different layers, acrylic
films~(200~$\mu$m thick) are inserted between individual fiber sheets in a
bundle.  Epoxy glue~(CY221/HY2967) was used to fabricate the bundle.  For
good optical contact with the IIT surface, we trimmed and polished the
surface of each bundle using a custom-made polishing machine.  The detector
geometry and construction logistics required us to perform bundle polishing
operations in the experimental hall, after installation of all the SciFi
modules, with very limited working space. A compact portable milling machine,
with dual carbide and diamond bits and pre-programmed automatic feed control,
was specially designed and built for this purpose\cite{PM-Mfg}. Following
polishing, bundles were inspected for flaws using a microscope system.

\subsection{Readout system}
\hspace{3mm} Our opto-electronics readout comprises two stages of
image-intensifiers, an optical lens and a CCD camera(See
Fig. \ref{fig:IIT-CCD}). The system was designed to ensure simplicity and
thus low cost.  The first stage is an electrostatic image-intensifier
(Hamamatsu V5502UX) with 100 mm diameter photocathode. The image is reduced
in diameter by a factor of 23 \% and the light yield is amplified by about 5. 
The bialkali photocathode has quantum efficiency 22\% at 430nm, which is the
peak wavelength in the emission spectrum of the scintillating fiber.  The
second stage is a micro-channel-plate(MCP) image-intensifier (Hamamatsu
MCP-IIT:V1366GX). The gain in light yield is about 1000 at a typical HV value
of 2600 volts.  We operate the MCP-IIT with pulsed HV to reduce unwanted
events.  The gate width is set to 100 $\mu$s, which corresponds to the decay
time of the fluorescent phosphor screen in the first stage IIT.  The optical
lens at the third stage reduces the image to 30 \% and projects it onto the
CCD camera (C3077). The CCD camera used here has 768 pixels and 493 pixels in
horizontal $(x)$ and vertical $(y)$ directions, respectively, with pixel size
11 $\mu$m $\times$ 13 $\mu$m. One fiber is seen by about $4 \times 4$ pixels.

\hspace{3mm} The video signal from the CCD is fed to a flash-ADC module
housed in a NIM crate, where $(x,y)$ positions of the CCD and the pulse
height are digitized. The digital outputs are then sent to 12 VME FIFO
modules. The readout time is about 30 msec.

\subsection{EL calibration} 
\hspace{3mm} To identify hit fibers from the CCD image, we have to know the
correspondence between fiber and CCD coordinates. For this purpose, we
illuminate selected fibers at regular intervals using an electro-luminescent
plate (EL). We will refer to these position calibrations as ``EL''
calibrations hereafter.

\begin{figure}[htbp]
	\begin{center}
	\epsfig{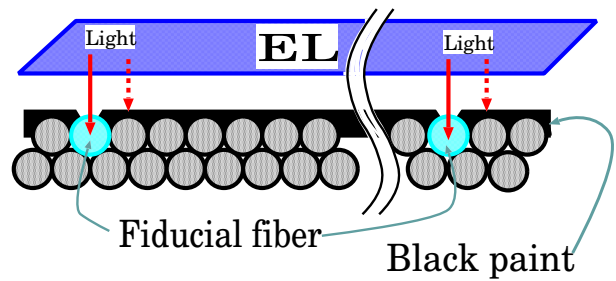}
	\caption[How to illuminate an EL]
	{
	\begin{minipage}[t]{12.0cm}
	\protect \small {
	Method for illuminating fiducial fibers with electroluminescent (EL)
plates. Black paint coating the fiber sheet is scraped off selected  
	fiducial fibers. Light from the EL plate enters only these fibers
(solid arrows), while the others remain 
	shielded (dashed arrow).}
	\end{minipage}
	}
	\label{fig:scraping1}
	\label{page:scraping1}
	\end{center}
\end{figure}

\hspace{3mm} We placed EL's at the edge of the fiber sheets (See
Fig. \ref{fig:bundle} in the previous subsection), where we coated the sheet
with black paint and then scraped the coating off selected fiducial fibers
(Fig. \ref{fig:scraping1}).  Thus, light from the EL goes only into the
fiducial fibers.  We illuminate only one out of every 10 or 20 fibers and
fibers at the edges of the sheets, to save labor and data size.  Positions of
the other fibers are interpolated from these fiducial measurements. We get
the fiber positions within 100 $\mu$m by this method. We measured the x, y,
and z positions of the edge fibers of each layer with accuracy of 1 mm during
installation, along with other fiducial marks on the fiber modules, to relate
fiber coordinates to the experiment coordinate system. Cosmic-ray muons were
used to obtain the final alignment information.

\section{Performance of the detector and analysis of the data}
\subsection{Hit definition and the pulse height}

\hspace{3mm} The energy loss of a minimum ionizing particle is estimated to
be 0.19 MeV in the fiber sheet and generates about 8 photoelectrons (p.e.) on
the IIT surface if the particle hits the fiber at the mid-point of SciFi
detector (2.4 m from the readout end), and 6.5 p.e. if it hits at the end of
the fiber (3.7 m from the readout).

\begin{figure}[htbp]
  \begin{center}
    \epsfig{file=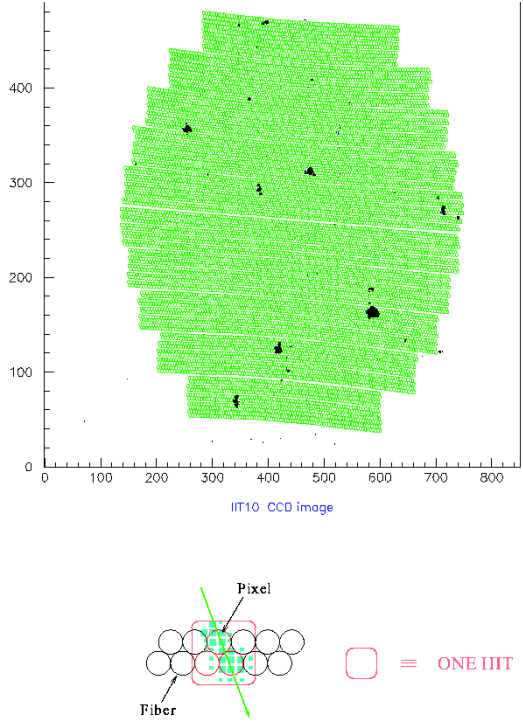,width=8cm}
    \caption{[Top] : Typical IIT image taken by CCD. The units on the axes
are CCD pixel number. The background of small circles denotes the
parameterized fiber positions determined from EL calibration. An enlarged
view of a hit is shown at bottom. The hit is defined as a cluster of fibers
which overlap the pixel clusters.}
    \label{fig:ccdimage}
  \end{center}
\end{figure}

\hspace{3mm} Fig. \ref{fig:ccdimage} shows a typical IIT image corresponding
to a particle track, read out by CCD: the dark spots represent hit CCD
pixels.  The fiber positions determined from EL calibration are also shown as
small circles. To reconstruct a hit, the hit finding algorithm starts
clustering CCD-pixels, finds all fibers which overlap the pixel cluster, and
clusters the fibers.  The hit is defined as a cluster of fibers which overlap
the pixel clusters: the hit position is calculated for the center of gravity
of the hit fibers (weighted by the number of overlapping pixels) in the fiber
cluster.

\hspace{3mm} The attenuation and reflection of photons inside the fiber was
measured during the quality test as described in Section
\ref{sec-quality}. The light yield is given by
\begin{equation}
  Y(d) = Y_0 (e^{-{d \over \lambda}} + R \cdot e^{-{2L-d \over \lambda}}),
  \label{att-eq}
\end{equation}
where $Y_0$ is the normalization constant, $d$ is the distance from the IIT
surface, $\lambda$ is the attenuation length (323 cm), $R$ is the
reflectivity of aluminum coated side (0.74), and $L$ is the length of fiber
(370 cm).

\begin{figure}[htbp]
  \begin{center}
    \epsfig{file=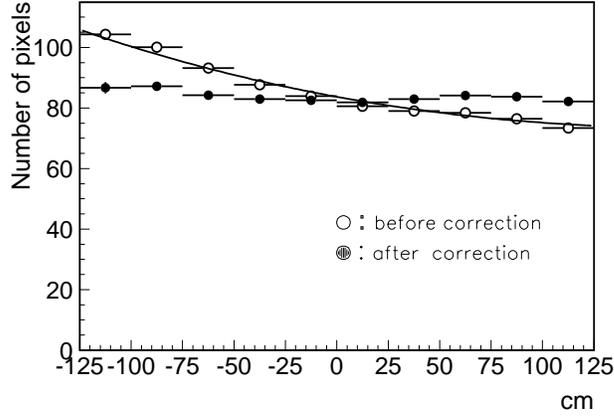,width=9cm}
    \caption{The mean values of $N_{pixel}$ as a function of hit position. 
    The data (open circles) are the measured values for cosmic-ray 
    muons. They agree well with the prediction from Eq. \ref{att-eq}(solid
    line). After the attenuation correction, the position dependence is
    negligible (solid circles).}
    \label{fig:attenuation}
  \end{center}
\end{figure}

\begin{figure}[htbp]
  \begin{center}
    \epsfig{file=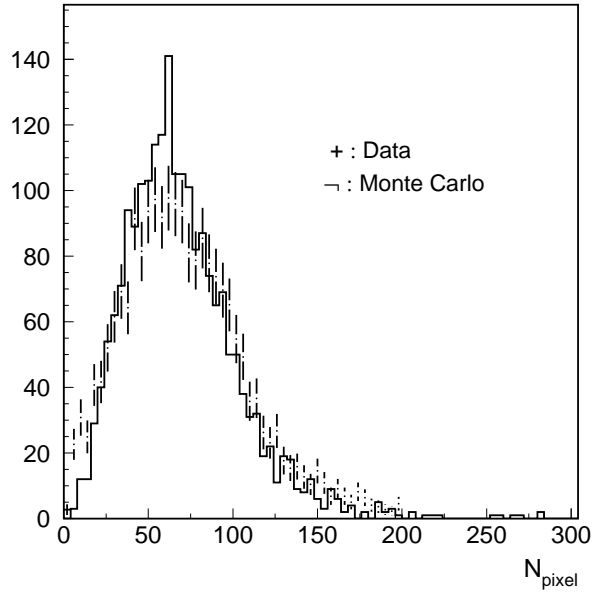,width=9cm}
    \caption{Typical distribution for the number of pixels ($N_{pixel}$) for
    cosmic-ray muons. Dots with error bars represent data and 
    the histogram shows pixel simulation results. We required $N_{pixel} \geq
    2$. } 
    \label{fig:Npixel}
  \end{center}
\end{figure}

\hspace{3mm} Fig. \ref{fig:attenuation} shows the mean values of the number
of pixels in a cluster ($N_{pixel}$) as a function of hit position. The IIT
surface is located at $x = d - 240$ cm (the center of detector is located at
$x = 0$). The data shown in open circles are the measured values for the
cosmic-ray muons and they agree well with the prediction by
Eq. (\ref{att-eq}) (solid line). After the attenuation correction, the
position dependence almost vanished, as shown in the figure (the solid
circles). The $N_{pixel}$ value for all track hits are normalized to that at
$x = 0$.
\hspace{3mm} Fig. \ref{fig:Npixel} shows the typical $N_{pixel}$ distribution
of data and a pixel simulation of cosmic-ray muons. $N_{pixel}$ is corrected
for the attenuation length and the incident angle.

\hspace{3mm} The response for a minimum ionizing particle (MIP) is calibrated
using cosmic-ray muons. The $N_{pixel}$ distribution for cosmic-ray muons
shows a peak at about 50 when it is normalized to the pulse height at $x =
0$. This means 8 p.e. corresponds to $N_{pixel} = 50$ at the peak position,
which agrees well with LED test data.

\subsection{SciFi Monte Carlo simulation - Pixel simulation } 
\hspace{3mm} The Monte Carlo simulation for the SciFi detector proceeds in
two steps. In the first, charged particles are tracked through the detector
using GEANT\cite{GEANT}. The energy loss of a particle (in GeV) is computed
whenever the particle traverses a fiber and is then converted to the number
of photoelectrons observed by the IIT/CCD imaging system, taking into account
attenuation along the fiber, reflection at the aluminium coated end and the
quantum efficiency of the IIT photocathodes.

\hspace{3mm} The photoelectrons are then converted to an image on the CCD
surface. Each photoelectron is offset from the center of the fiber according
to a gaussian distribution measured using single photoelectrons during LED
testing of the IITs. The size of the pixel cluster corresponding to a single
photoelectron is then chosen from distributions parameterized from the LED
data and the pixel cluster is built assuming circular symmetry. Finally, the
tuning of the energy scale (i.e. how many photoelectrons correspond to the
energy deposition from a minimum ionising particle) is carried out using
cosmic-ray muons.

\subsection{Track and vertex reconstruction}
\hspace{3mm} The track and vertex finding algorithm proceeds in 4 steps.
First, the track finder searches for 2-dimensional (2D) track candidates (in
the XZ or YZ planes) which hit at least 3 layers of SciFi modules. A 3D track
is reconstructed by finding the best combination of two X-view and Y-view 2D
tracks, comparing the starting and ending points and checking the overlap of
the 2D tracks. A vertex is defined as an intersection of 3D tracks, and the
vertex position is determined with full 3D $\chi^2$ minimization fit.  If
there is only one track in the event, the mid-point of the nearest upstream
water tube is taken as the event vertex. Finally, using the reconstructed
vertex position, the track finder starts searching for extra tracks near the
vertex, which are mainly short tracks.  The last step is very important
because most proton tracks from charge current (CC) interactions are short
and may not be found in the first iteration of track finding.

\begin{figure}[htbp]
  \begin{center}
    \epsfig{file=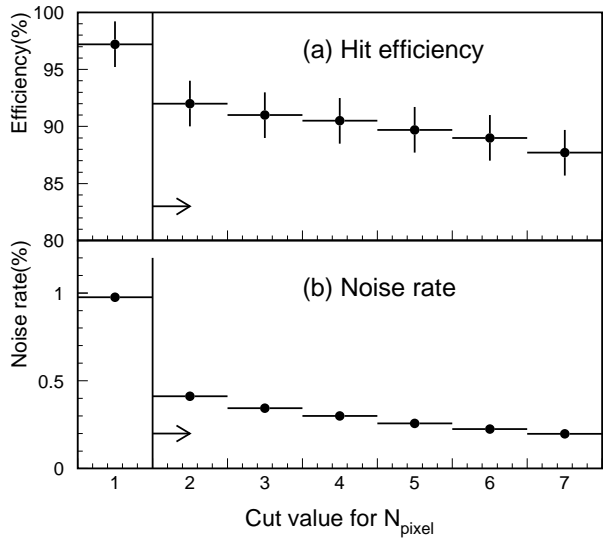,width=9cm}
    \caption{(a) Hit efficiency and (b) noise rate for various cut values on
     $N_{pixel}$, estimated with cosmic-ray data. The arrows show the chosen
     analysis cut.}
    \label{fig:npixcut}
  \end{center}
\end{figure}

\begin{figure}[htbp]
  \begin{center}
    \epsfig{file=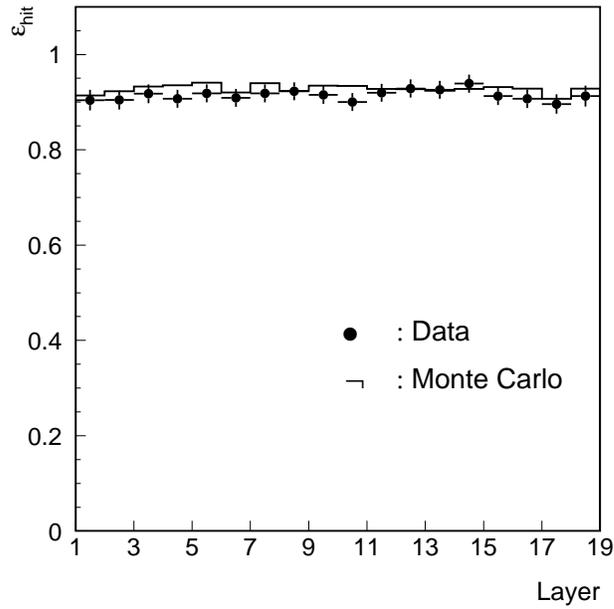,width=9cm}
    \caption{Hit efficiency for each layer estimated with cosmic-ray 
     muons(solid circles for data and histogram for Monte Carlo simulation).}
    \label{fig:hiteff}
  \end{center}
\end{figure}

\begin{figure}[htbp]
  \begin{center}
    \epsfig{file=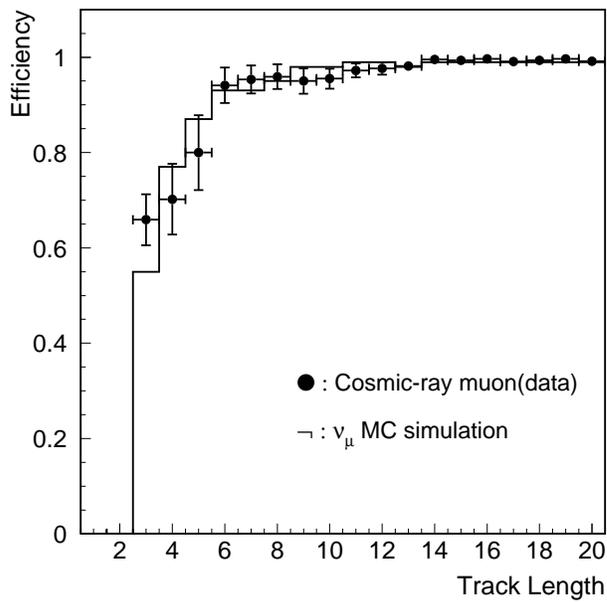,width=9cm}
    \caption{Track finding efficiency as a function of track length,
    expressed  in terms of the number of SciFi layers traversed, for
    cosmic-ray muons (solid circles) and for muon tracks from  $\nu_\mu$
    charged-current interactions generated by the Monte Carlo simulation
    (histogram).}  
    \label{fig:trkeff}
  \end{center}
\end{figure}

\hspace{3mm} The efficiency of track finding strongly depends on hit
efficiency and the density of noise hits in the events. Hit efficiency
($\varepsilon_{hit}$) was evaluated using cosmic-ray muons which penetrate
whole SciFi modules as
\begin{equation}
\varepsilon_{hit} =  {{Number \; of \; hits \; on \; the \; track }
\over {Number \; of \; expected \; hits \; on \; the \; track}} .
\end{equation}
In order to keep the hit efficiency reasonably high and to keep the noise
rate as low as 0.5 \%, we required a minimum value of $N_{pixel}$ for a
hit. Fig. \ref{fig:npixcut} shows the hit efficiency and the noise rate at
various cut values on $N_{pixel}$. Based on this figure, we decided to
require $N_{pixel} \geq 2$.  As a result, the hit efficiency was found to be
92$\pm$2 \% on average, with almost no noise contamination ($<$0.5
\%)\footnote{After upgrading the HV system of IIT in Summer, 1999, we found
the hit efficiency to be 96 $\pm$ 2 \% with the same noise
level.}. Fig. \ref{fig:hiteff} shows the hit efficiency for each layer.
\hspace{3mm} Fig. \ref{fig:trkeff} shows the track finding efficiency, which
was estimated using cosmic-ray muons and muon tracks from $\nu_\mu$ CC
interactions generated in the Monte Carlo simulation. They agree with each
other: 98$\pm$2 \% for long tracks (traversing more than 7 SciFi layers) and
85$\pm$6 \% for short tracks. Here, the horizontal axis is the number of
layers which the muon passes through.

\hspace{3mm} Vertex resolution estimated using Monte-Carlo events($\nu_\mu$
CC events) was 1.5 mm in the beam direction and 1.5 mm perpendicular to the
beam.

\begin{figure}[htbp]
  \begin{center}
    \epsfig{file=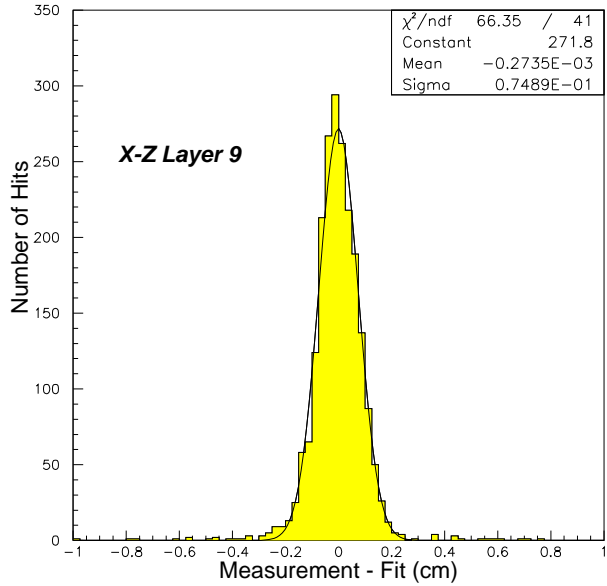,width=9cm}
    \caption{Typical residual distribution for cosmic-ray muon track fits 
    for a representative layer (layer 9). 
    To obtain this figure, we excluded the measured point in layer 9, 
    and tested it against the fit results.}
    \label{fig:residuals}
  \end{center}
\end{figure}

\hspace{3mm} The position resolution of SciFi detector was estimated from the
residual distribution for hits. Since multiple Coulomb scattering effects are
not negligible, we implemented a Kalman filtering/smoothing
algorithm\cite{kalman} in order to take multiple scattering into account
during the track fit. Fig. \ref{fig:residuals} shows a residual distribution
of hits for cosmic-ray muons. The figure shows that hit position resolution
is 0.8 mm.\footnote{Using a simple straight-line fit instead of Kalman Filter
method, the rms of the corresponding residual distribution was 1.5 mm.} At
present, performance is limited by multiple scattering effects in the water
tank and imperfect alignment of the sheets. If we can correct for the
detailed alignment parameters, such as the curvature and rotation of the
sheets, we expect the hit position resolution to improve to about 0.5mm,
which is dominated by multiple scattering effects.

\subsection{Neutrino event reconstruction}
\begin{figure}[htbp]
  \begin{center}
    \epsfig{file=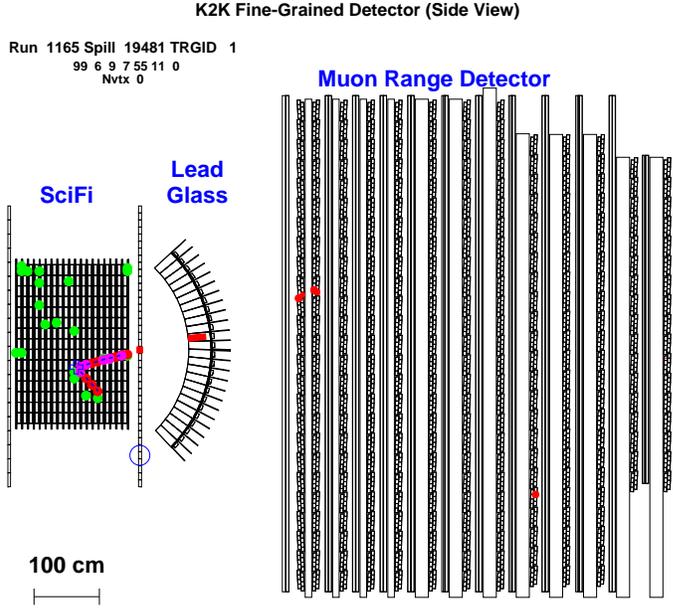,width=9cm}
    \caption{Typical $\nu_\mu$ event candidate in the Fine-Grain detector
    including SciFi detector.}
        \label{fig:numu}
  \end{center}
\end{figure}

\hspace{3mm} Fig. \ref{fig:numu} shows a typical $\nu_\mu$ CC quasi-elastic
(qe) interaction ($\nu_\mu + n \rightarrow \mu^- + p$) candidate. The longer
track is a muon, which hits the scintillating counters and lead glass
counters and reaches the muon range detector, and the shorter track is a
proton.

\begin{figure}[htbp]
  \begin{center}
    \epsfig{file=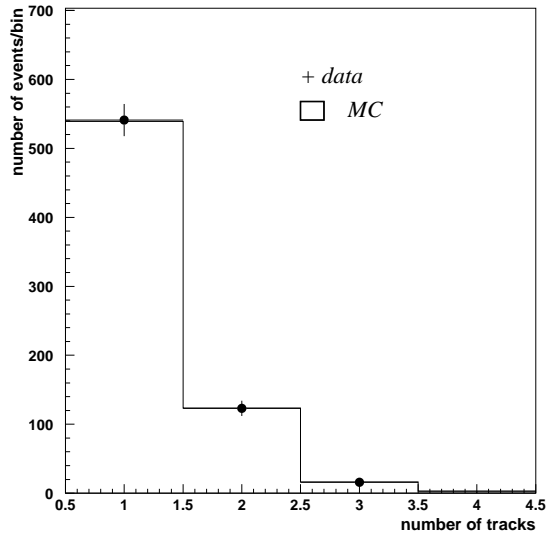,width=9cm}
    \caption{Number of tracks found in $\nu_\mu$ event candidates. The total
    number of events in a Monte Carlo prediction is normalized to that in the
    data. The errors shown are statistical only. The systematic error at
    each data point is about 5 \%. } 
    \label{fig:ntrack}
  \end{center}
\end{figure}

\hspace{3mm} Fig. \ref{fig:ntrack} shows the distribution of the number of
the reconstructed tracks from the $\nu_\mu$ interaction vertex in $\nu_\mu$
event candidates. The total number of events in a Monte Carlo prediction is
normalized to that in the data. It shows that charged particles in
$\nu_{\mu}$ CC events are succesfully reconstructed and the data (solid
circles) are well reproduced by Monte Carlo prediction (histogram). We
estimated the fraction of tracks with misreconstructed trajectories to be
less than 4 \%. 

\subsection{Particle identification}
\hspace{3mm} Although tracking is the main purpose of the SciFi detector, we
may use it for particle identification. Electrons can be identified from
SciFi hit patterns, and are useful for identifying the $\nu_e$ events. The
energy of an electron can be estimated from the number of hits in the shower,
{\it i.e.} $E_e$ = $\sim$ 10 MeV/hit for SciFi detector, which is estimated
from a Monte Carlo study and a beam test with a prototype detector. The
energy resolution of electrons was measured to be about 28 \% for 0.3 GeV in
the beam test and it is estimated to be 15 \% for 1.0 GeV electrons.

\begin{figure}[htbp]
  \begin{center}
    \epsfig{file=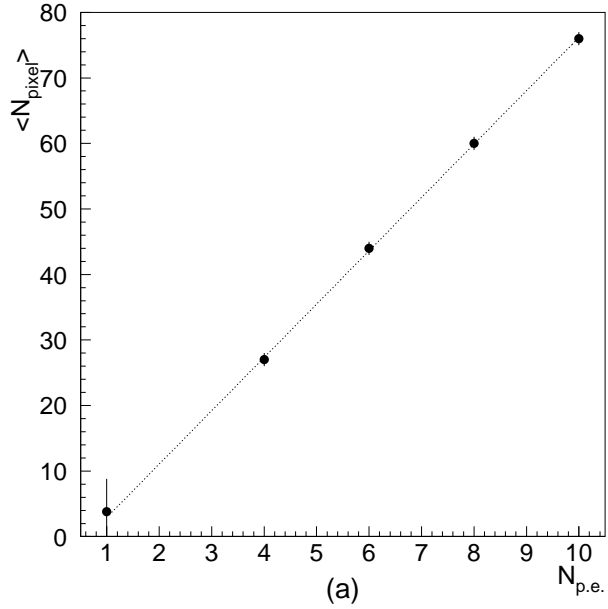,width=9cm}
    \epsfig{file=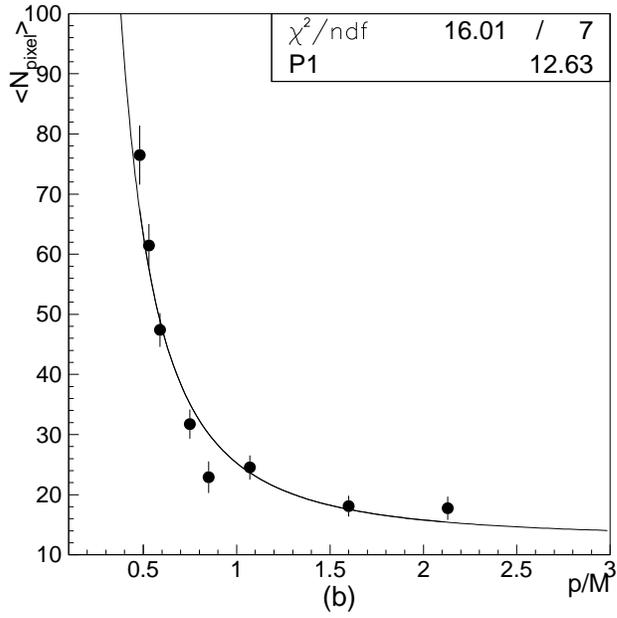,width=9cm}
    \caption{Linearity of IIT : (a) The mean value of $N_{pixel}$ as a 
    function of $N_{p.e.}$(the number of photo-electrons) from the LED test.
    (b) The mean value of $N_{pixel}$ as a function of $p/M$ from the beam 
    test with a proto-type detector using the proton and pion beams. The
    simplified Bethe-Bloch formula (solid curve) reproduces the data points
    well.} 
    \label{fig:Linearity}
  \end{center}
\end{figure}

\begin{figure}[htbp]
  \begin{center}
    \epsfig{file=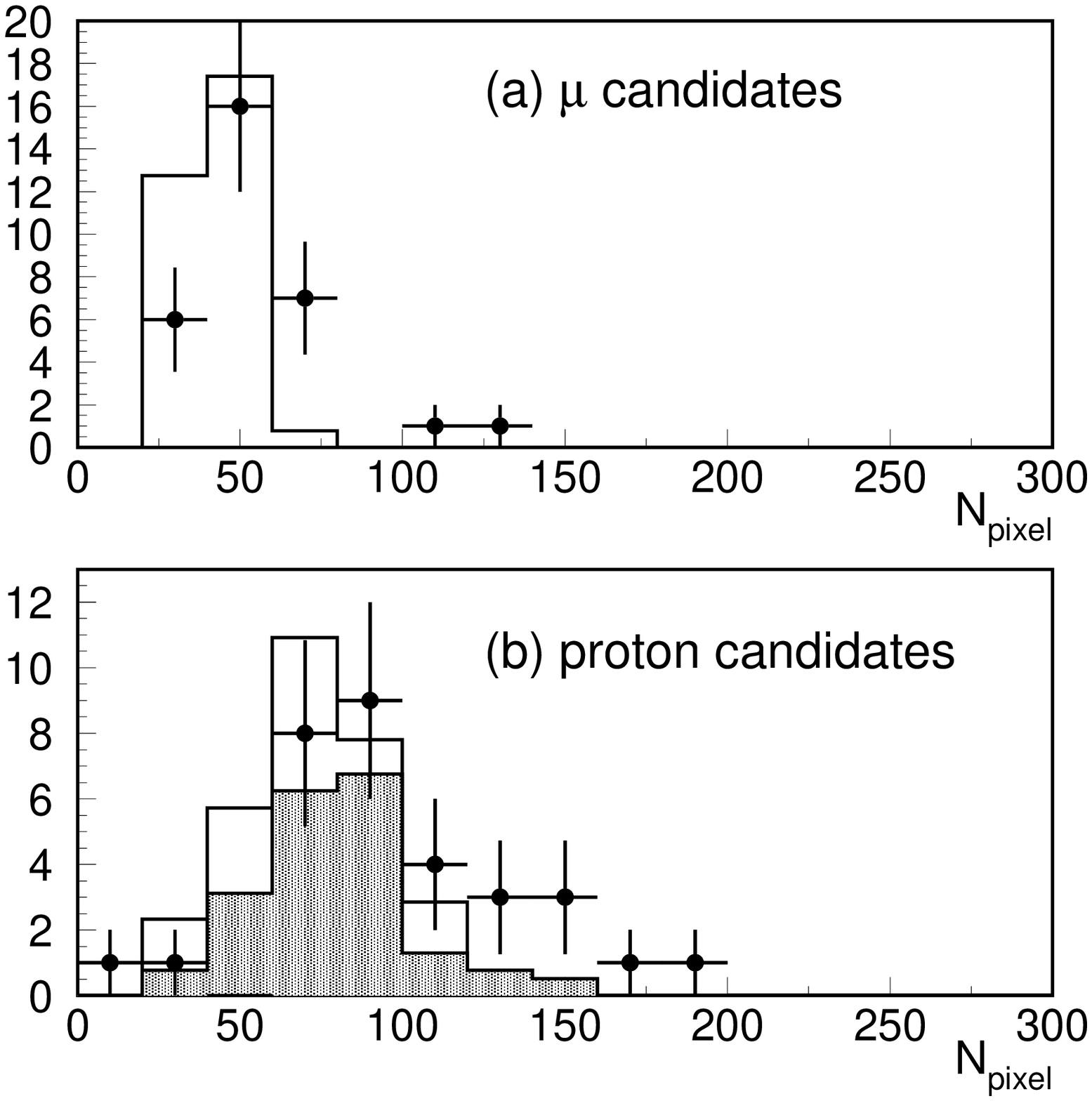,width=9cm}
    \caption{The $N_{pixel}$ distribution for (a) $\mu$ candidates and (b)
    proton candidates from $\nu_{\mu}$ events containing two tracks.
    The data(solid circles) agree with 
    the Monte Carlo prediction(histogram). 
    The hatched histogram shows the proton contribution.}
    \label{fig:pid}
  \end{center}
\end{figure}

\hspace{3mm} From LED runs and the beam test, the number of pixels
($N_{pixel}$) shows good linearity with the number of photoelectrons
($N_{p.e.}$). Fig. \ref{fig:Linearity}(a) shows the correlation of
$<N_{pixel}>$ and $N_{p.e.}$ from the LED test.  Fig. \ref{fig:Linearity}(b)
shows the results of the beam test of the proton and pion beams. The
horizontal axis is the $p/M$ of the particle and the vertical axis is the
mean value of $N_{pixel}$. The simplified Bethe-Bloch formula reproduces the
data points well.
\hspace{3mm} $N_{pixel}$ of the tracks may be used to separate protons from
muons or pions. A feasibility study was performed using $\nu_\mu$ CCqe event
candidates, which contain one muon track and one proton
track. Fig. \ref{fig:pid} shows $N_{pixel}$ distribution for (a) muon and (b)
proton candidates. The figure shows that the proton candidates have larger
$N_{pixel}$ values because protons will lose more energy in the fiber sheet.

\section{Summary}
\hspace{3mm} A very large scintillating fiber (SciFi) tracking detector for
the K2K long baseline neutrino oscillation experiment has been in operation
since March, 1999. Track finding efficiency is 98$\pm$2 \% for long muon
tracks (those which intersect more than 5 fiber planes), and 85$\pm$6 \% for
short tracks, which were estimated using cosmic-ray muons and a Monte Carlo
simulation. The position resolution per layer is about 0.8 mm. The SciFi
detector has demonstrated its capability for reconstruction of $\nu_\mu$
interactions. The pulse heights for cosmic-ray muons have been stable within
10 \% after one year of operation.  In addition, the SciFi detector offers
the possibility of performing electron and proton identification.

\section{Acknowledgements}
\hspace{3mm} We thank Mr. S. Ishikawa and Mr. T. Kawai of Nagoya University
for designing and constructing the fiber spooling machine. We also thank
Prof. K. Niwa and his group, especially Dr. T. Nakano, of Nagoya University
for their cooperation. Thanks are also due to Mr.  C. Lindenmeyer,
Mr. P. Mulligan, and Mr. J. Roze, for the design, construction, and operation
of the fiber bundle polishing machine.  We also thank M. Powlowski, T. Raza,
and T. Kato (SUNY) for helping the fiber sheet assembly.

\end{document}